\documentclass[showpacs,preprintnumbers,amsmath,amssymb]{revtex4}
\usepackage{amsfonts}
\usepackage{amssymb}
\usepackage{graphics} 
\usepackage{epsfig}
\usepackage{fancyheadings}
\usepackage{url}
\usepackage{multirow}
\newcommand {\be}{\begin{equation}}
\newcommand {\ee}{\end{equation}}
\newcommand {\ba}{\begin{eqnarray}}
\newcommand {\ea}{\end{eqnarray}}
\newcommand {\ra}{\rightarrow}
\newcommand {\tanb}{tan$\beta~$}

\begin{document}

\title{Observability of Light Charged Higgs Decay to Muon in Top Quark Pair Events at LHC}
\author{M. Hashemi}
\affiliation{Physics Department and Biruni Observatory, Shiraz University, Shiraz 71454, Iran}
\affiliation{Institute for Research in Fundamental Sciences (IPM),P.O. Box 19395-5746, Tehran, Iran}

\begin{abstract}
In this paper the charged Higgs signal through the decay to a pair of muon and neutrino ($H^{\pm}\ra \mu\nu$) is analyzed. 
The analysis attempts to estimate the amount of muonic signal of the charged Higgs at LHC at a center of mass energy of 14 TeV. 
The signal process is the top quark pair production with one of the top quarks decaying to a charged Higgs (non SM anomalous top decay) and the other decaying to a W boson which is assumed to decay hadronically to two light jets. 
Due to the small branching ratio of charged Higgs decay to muon, 
results are quoted for data corresponding to an integrated luminosity of 300 $fb^{-1}$ 
which is expected to be collected at the LHC high luminosity regime. It is shown that a signal 
significance close to 5$\sigma$ down to below 1$\sigma$ is achievable for a charged Higgs mass in the range 80 GeV $< m(H^{\pm}) <$ 150 GeV taking the top quark pair production with both top quarks decaying to W bosons as the main irreducible background.

\end{abstract}

\maketitle

\section{Introduction}

The Large Hadron Collider (LHC) at CERN is one of the biggest high energy physics experiments which is 
expected to provide fruitful information about the possible new phenomena at the scale of sub-atomic 
collision events. Currently running at a center of mass energy of 7 TeV, LHC has delivered almost $5 fb^{-1}$ data to each one of the CMS and ATLAS experiments. 
Obviously one of the exciting searches is the search for the Higgs boson which is expected to be responsible for the mass of particles through the Higgs mechanism \cite{Higgs1,Higgs2,Higgs3,Higgs4,Higgs5}. The current results of Higgs boson searches in CMS exclude the Higgs boson mass in the range 127-600 GeV at 95$\%$ confidence level \cite{SMHiggsCMS}, while the excluded ranges by ATLAS are 112.7-115.5 GeV, 131-237 GeV and 251-468 GeV at 95$\%$ confidence level \cite{SMHiggsATLAS}. Both above results obviously confirm and extend the current Tevatron results \cite{TevatronHiggs}.\\
The above news imply that the next few years of LHC experiments will be very exciting and provide a variety 
of information which develop our knowledge and level of understanding of the Standard Model and in 
case of observing the Higgs boson a reasonable understanding of the origin of mass spectrum through the Higgs mechanism is achieved.\\
While an extensive effort is ongoing at the Tevatron and LHC experiments in search for the Standard Model Higgs boson, there are arguments that a single Higgs boson predicted through the SM Higgs mechanism is not a complete theory. In fact the problem of quadratic divergence of the Higgs boson mass when quantum corrections are included does not receive a solution as elegant as the Supersymmetry (SUSY) which introduces a super-partner for each particle whose spin is different by half unit \cite{martin}. Making a double universe this way, each fermion acquires a boson as a partner and vice versa. It is shown that within these models the quadratic divergence of the Higgs boson mass reduces to a logarithmic function of the cut-off of the theory. \\
One of the interesting results of supersymmetric models is that instead of a single Higgs boson, several such particles with different properties are produced. In the minimal supersymmetric extension to Standard Model, the so called MSSM, five physical Higgs bosons are predicted. Among these five, two are charged and the other three are neutral.\\
It might look difficult at the beginning to distinguish between the SM neutral Higgs boson and a neutral MSSM Higgs boson. However the decay preferences of the SM Higgs boson and MSSM neutral Higgs bosons are different. i.e. there are different branching ratio of decays when SM Higgs and MSSM neutral Higgs bosons are compared and therefore different strategies are taken in the search for them \cite{ATLASTDR,CMSTDR}. \\
A long time attempt has been dedicated to the search for the MSSM neutral Higgs bosons in the past and current experiments. The LEP results on such searches are reported in \cite{LEPNeutralMSSM} where the lower limits on the mass of neutral Higgs bosons are found to be $m_{h^{0}}>91.0$ GeV and $m_{A^{0}}>91.9$ GeV and low tan$\beta$ values,
 i.e. $0.5 < $tan$\beta <2.4$ are excluded. The Tevatron results are also quoted in \cite{TevNeutralMSSM} where high tan$\beta$ values are excluded.
The LHC collaborations have also started an intensive search for such neutral MSSM Higgs bosons. The CMS Collaboration has recently published the first results of such searches in \cite{H2tautau} which has set the strongest limits on tan$\beta$. \\
Another direction of search is looking for the charged Higgs. Being a charged particle, it provides unique signatures which makes it distinguishable from neutral Higgs bosons. The decay channels are different and due to being a spinless particle, it produces different topology and kinematics of events when compared to the electroweak gauge bosons such as $W^{\pm}$. Observation of such a different particle is a crutial signature of models beyond Standard Model such as MSSM.
This is why a special care has been taken in searches for this particle. \\
The LEP search for the charged Higgs has been reported in \cite{LEPCH} where a charged Higgs mass below 78.6 GeV is excluded at 95$\%$ confidence level. This limit is of course softer than that obtained in indirect searches in \cite{LEPNeutralMSSM} where a charged Higgs mass below 125 GeV is excluded at high tan$\beta$ values (tan$\beta \gtrsim 10$). The Tevatron searches are reported in \cite{d01,d02,d03,d04} by the D0 Collaboration and \cite{cdf1,cdf2,cdf3,cdf4} by CDF Collaboration. The overall result is $2 <$ tan$\beta < 30$ for $m(H^{\pm}) > 80 $ GeV while wider $\tan \beta$ values are available for higher charged Higgs masses. It should be noted that different strategies are used in the search for this particle. In addition to direct search limits and limits obtained from searches for the neutral MSSM Higgs bosons, indirect search limits are obtained using analyses which look for any deviation in the measured SM cross sections as a hint for the existence of beyond SM particles such as the charged Higgs bosons. A review of direct and indirect searches at the Tevatron can be found in \cite{tevrev}.\\
Results from B-factories are also used to put some constraints on the charged Higgs mass as the existence of a charged Higgs in B-meson decay diagrams may alter branching ratio of decays and can be verified experimentally. A study of the $b\ra s\gamma$ transition process using CLEO data has excluded a charged Higgs mass below 295 GeV at 95 $\%$ C.L. in 2HDM Type II with \tanb higher than 2 in \cite{cleo}. This limit is obviously stronger than direct search limits, however, there have been studies such as the one reported in \cite{weaken} where it has been shown that in a general 2HDM with complex Higgs-fermion coupling, the charged Higgs mass can be as low as 100 GeV.  
The above direct searches carried out at LEP, Tevatron and LHC are based on the charged Higgs decay to tau and the tau neutrino, i.e. $H^{+} \ra \tau^{+}\nu_{\tau}$.
This is because the light charged Higgs decays predominantly to the tau lepton and its signature is the excess of tau leptons over the Standard Model expectation which is also inferred as some imbalance in the number of leptons from W decays. This is the so called lepton universality breaking phenomenon. The charged Higgs is produced from an on-shell (off-shell) top quark if it is lighter (heavier) than the top quark. \\
The $\tau$ lepton identification algorithm follows theoretical and phenomenological studies of different decay channels of this particle. These studies have improved the $\tau$ lepton identification over the years. An example of a comprehensive theoretical study of the $\tau$ lepton decay channels can be found in \cite{tau1}. In \cite{tau2} and \cite{tau3} algorithms for the $\tau$ lepton identification were proposed for Tevatron and LHC. In a different work an update of the algorithm specific to the case of heavy charged Higgs was proposed in \cite{tau4}. The charged Higgs analyses searching for the $\tau\nu$ decay mode extensively use the idea proposed in above references.\\
The main process for the light charged Higgs is $t\bar{t}$ production with at least one of the top quarks decaying to a charged Higgs, i.e. $t\bar{t} \ra H^{\pm}W^{\mp}b\bar{b}$. This is to be compared with a background from $t\bar{t} \ra W^{\pm}W^{\mp}b\bar{b}$. Therefore the top decay to charged Higgs, i.e. $t \ra H^{+}b$ is compared with $t \ra W^{+}b$ and 
since W bosons equally decay to electron, muons and taus, the existence of a charged Higgs in the top decay is concluded by observing an excess of tau leptons over what we expect from $t \ra W^{+}b$. \\
A helping character of top decay to charged Higgs is that the mass of the charged Higgs is taken to be higher than the W boson mass required by the direct search results of LEP reported in \cite{LEPCH}. Therefore a more energetic tau lepton produced in charged Higgs decay is better identified than its partner from W boson decay. The same arguments apply to a heavy charged Higgs which is produced in gluon-gluon interactions producing a top quark in association with the charged Higgs, i.e. $gg \ra t\bar{b}H^{-}$. The details of this process can be found in \cite{ggtbh1,ggtbh2}. Again for the heavy charged Higgs, the main background is taken to be the Standard Model process $t\bar{t} \ra W^{+}W^{-}b\bar{b}$ and the same approach as above is used for the search.\\
In CMS and ATLAS collaborations, several Monte Carlo studies have been carried out for the light and heavy charged Higgs. The CMS studies of the light and heavy charged Higgs decay to tau are documented in \cite{CMSLCH,CMSHCH}. The heavy charged Higgs decay to top and bottom quarks has also been studied in CMS in \cite{CMSHCHTB}. Although the latter does not lead to a significant signal due to the large hadronic background, the analyses in \cite{CMSLCH,CMSHCH} clearly show that a large area of parameter space in terms of $m(H^{\pm})$ and tan$\beta$ is available for observation of this particle. The ATLAS collaboration also quotes their search results on the potential of observing a charged Higgs boson in different final states in \cite{ATLASTDR}.\\
An overall conclusion by CMS and ATLAS collaborations is that the charged Higgs decay to $\tau\nu$ will be the most promising decay channel in the search for this particle and will certainly be taken as the highest priority channel for the early data analysis at LHC. That is why no more attention has been paid to the charged Higgs rare decays such as decay to muon in the current searches. Since $H^{\pm} \ra \tau\nu$ provides almost enough statistics for a wide range of the charged Higgs mass, any possible discovery is expected to happen in $H^{\pm} \ra \tau\nu$ channel. \\ 
In this paper the attempt is to verify the observability of charged Higgs decay to muon quantitatively. The aim is to verify whether this signal would be finally observable at the LHC high luminosity regime or not. To the best of our knowledge, no detailed analysis has been published in the literature focusing on this channel. Obviously this channel is not suitable as an early data analysis and can not be considered as a discovery channel because if a charged Higgs exists, the tauonic decay signal would have much more statistics to tell us about the existence of the charged Higgs particle. However by the time LHC will have provided enough data to study the rare decays of this particle, the muonic signal would be a good candidate providing a clean signature for studies of the charged Higgs boson spin, coupling to fermions, etc. For instance, studies of the muon flight direction and $p_{T}$ distribution could provide opportunity to verify the charged Higgs spin through the shape fits in a similar way as has been done for $W^{\pm}$ bosons \cite{Watlas}. A detailed study of such phenomena could be carried out by LHC experiments when enough data is available. \\
As a comparison, the charged Higgs decay to muon suffers from a small coupling which is proportional to the lepton mass and the branching ratio in this case is roughly two orders of magnitude smaller than the case of tauonic decay. This is compensated to a reasonable extent by the fact that muons are identified with a high efficiency thanks to the good detector functionality and the powerful muon detection systems of LHC experiments. On the contrary, the tau lepton identification requires a sequence of selection cuts and the total efficiency is low \cite{tauid}. Therefore the charged Higgs signal in this case is reduced significantly from what is produced at the beginning although still being higher than that of charged Higgs decay to muon. The decay to $c\bar{b}$, $c\bar{s}$ or $u\bar{s}$ does not lead to a promising signal as they all suffer from the large hadronic background from QCD multi-jet events.\\
Above thoughts motives a study of the charged Higgs decay to muon in a quantitative way and estimation of the signal significance as a function of the charged Higgs mass. Since the light charged Higgs comes mainly from the $t\bar{t}$ production which has a high cross section at LHC, this study focuses on the light charged Higgs which is expected to give the highest sensitivity. Extrapolation to the case of heavy charged Higgs can be made later or in a separated study. Therefore this is a study with the goal of evaluating the possibility of observing $H^{\pm} \ra \mu\nu$ in the region of light charged Higgs, i.e. below the top quark threshold.

\section{Signal and Background identification and simulation} 
\begin{figure}
\begin{center}
\includegraphics[width=0.6\textwidth]{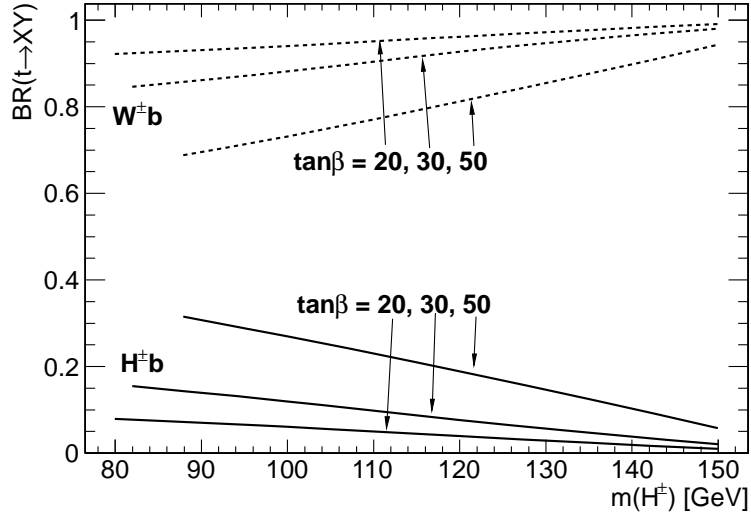}
\end{center}
\caption{Branching ratio of top decay to charged Higgs and W boson as a function of its mass.}
\label{brt2xy}
\end{figure}

\begin{figure}
\begin{center}
\includegraphics[width=0.6\textwidth]{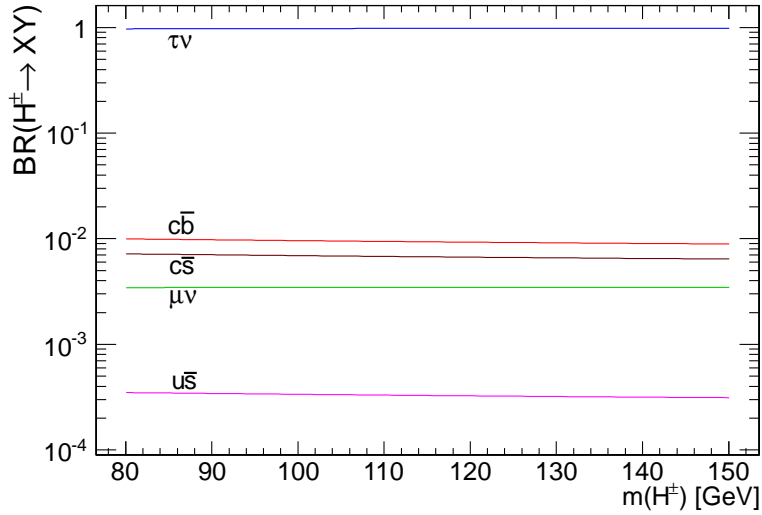}
\end{center}
\caption{Branching ratio of charged Higgs decays a function of its mass.}
\label{brh2xy}
\end{figure}

The signal for this analysis is $t\bar{t} \ra H^{\pm}W^{\mp}b\bar{b}$ followed by $H^{\pm} \ra \mu \nu$ and $W^{\mp} \ra j j$ where only the hadronic decay of the W boson is considered to be able to reconstruct the Higgs boson transverse mass. Therefore the final state consists of a muon, four jets two of which are b-jets, and some amount of missing transverse energy originating from the neutrino in the event. Figure \ref{brt2xy} shows the top quark branching ratio of decay to W and charged Higgs as a function of the charged Higgs mass. The charged Higgs masses under study in this work start from the direct search lower limit in \cite{LEPCH} and cover a mass range up to 150 GeV where the signal cross section becomes very small. The tan$\beta$ used in the study is 20, as higher values are close to or inside the excluded area in \cite{H2tautau}. The charged Higgs branching ratio of decays are shown in Fig. \ref{brh2xy}. As is observed the decay to muon has a branching ratio of roughly $0.35 \times 10^{-2}$. The lighter charged Higgs, the higher branching ratio of top decay to charged Higgs. However it is below 0.1 for all the region of study and when multiplied by the charged Higgs branching ratio of decay to muon results in a small cross section compared to background. \\
The main background is $t\bar{t} \ra W^{\pm}W^{\mp} b\bar{b}$ followed by $W^{\pm} \ra \mu \nu$ and $W^{\mp} \ra j j$. Table \ref{xsec} lists the signal and background cross sections multiplied by branching ratios. In calculating the numbers quoted in Tab. \ref{xsec}, MCFM 5.7 \cite{mcfm} has been used with the parton distribution function (PDF) set to MRST 2006 which was taken by linking LHAPDF 5.8.3 \cite{lhapdf} to MCFM. The top quark mass is set to 175 GeV. The total $t\bar{t}$ cross section using the above PDF set is 879 pb. The W boson branching ratios are taken from the Particle Data Group (PDG) data \cite{pdg} (BR($W\ra j j) \simeq$ 0.68 and BR($W\ra \mu\nu) \simeq$ 0.106). The charged Higgs branching ratios are calculated using HDECAY 3.51 \cite{hdecay}. The event simulation and calculation of branching ratios of top quark decays are done using PYTHIA 6.4.21 \cite{pythia}. When generating events the same PDF set is linked to PYTHIA through the LHAPDF interface. The study is based on a parton level generation of events followed by hadronization and jet reconstruction using the PYTHIA jet reconstruction tool PYCELL. The initial and final state QCD and QED radiations are also taken into account in the simulation settings. The jet reconstruction cone opening angle is 0.5 and only jets with pseudorapidity $|\eta|<3$ are reconstructed. Here the pseudorapidity is defined as $\eta=-ln(\tan(\theta/2))$ where $\theta$ is the polar angle with respect to the beam axis. The MSSM, $m_{h}$-max scenario is used with the following parameters: $M_{2}=200$ GeV, $M_{\tilde{g}}=800$ GeV, $\mu=200$ GeV and $M_{SUSY}=1$ TeV.

\begin{table}
\begin{center}
\begin{tabular}{|c|c|c|c|c|c|}
\hline
& \multicolumn{4}{|c|}{Signal} & Background \\ 
\hline 
$m_{(H^{\pm})}$ & 80 GeV  & 100 GeV & 120 GeV & 150 GeV & - \\
\hline
$\sigma \times $BR & 302 fb & 228 fb & 149 fb & 38 fb & 124 pb\\
\hline
\end{tabular}
\end{center}
\caption{Signal and background cross sections times branching ratios. For the signal tan$\beta$=20 has been used. \label{xsec}}
\end{table}

\section{Event Analysis and Selection}
\begin{figure}
\begin{center}
\includegraphics[width=0.6\textwidth]{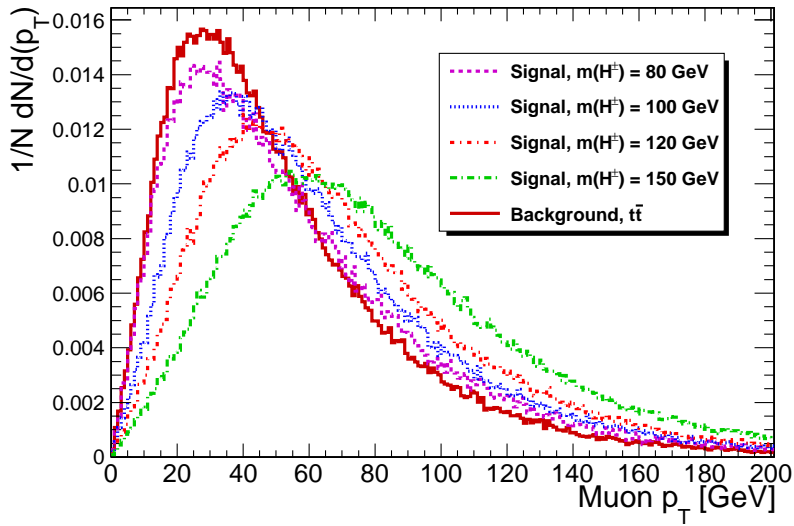}
\end{center}
\caption{Muon transverse momentum distributions in signal events with different charged Higgs massed and the background sample.}
\label{muonpt}
\end{figure}

\begin{figure}
\begin{center}
\includegraphics[width=0.6\textwidth]{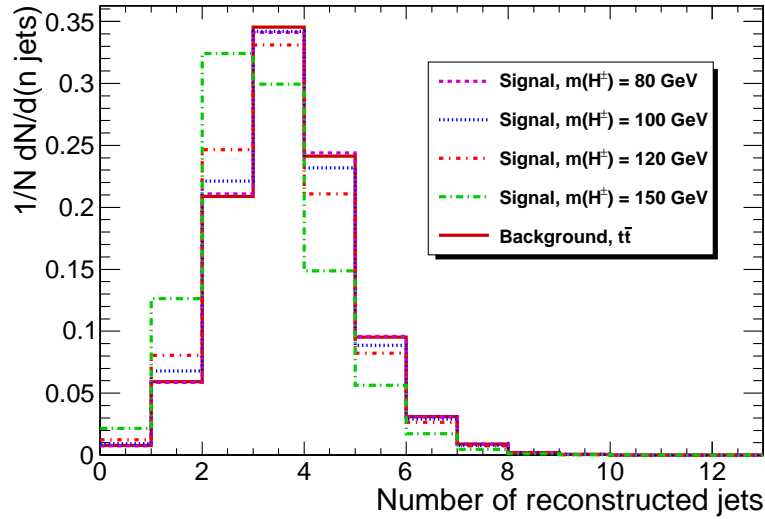}
\end{center}
\caption{Number of reconstructed jets passing $E_{T}>30$ GeV and $|\eta| < 2.5$ in signal and background samples.}
\label{njet}
\end{figure}

\begin{figure}
\begin{center}
\includegraphics[width=0.6\textwidth]{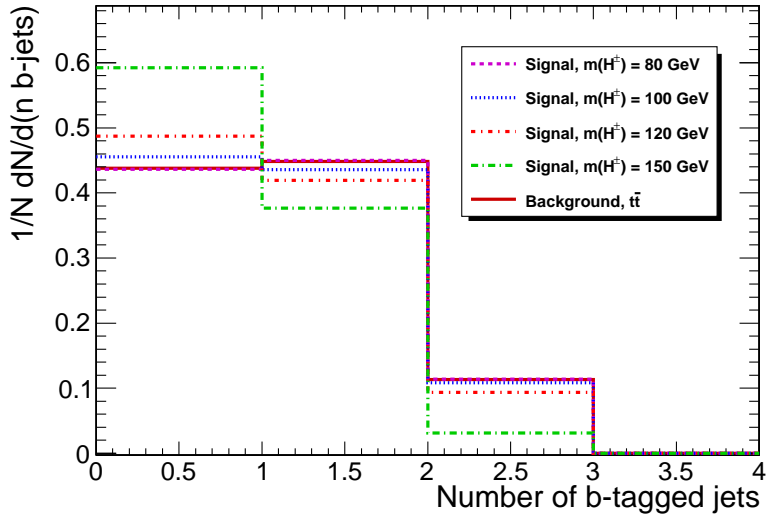}
\end{center}
\caption{Number of b-tagged jets in signal and background events. The b-tagging algorithm is based on matching jets which pass the kinematic cuts, with the parton level b-quarks. A b-tagging efficiency of 50$\%$ has been used.}
\label{nbjet}
\end{figure}

\begin{figure}
\begin{center}
\includegraphics[width=0.6\textwidth]{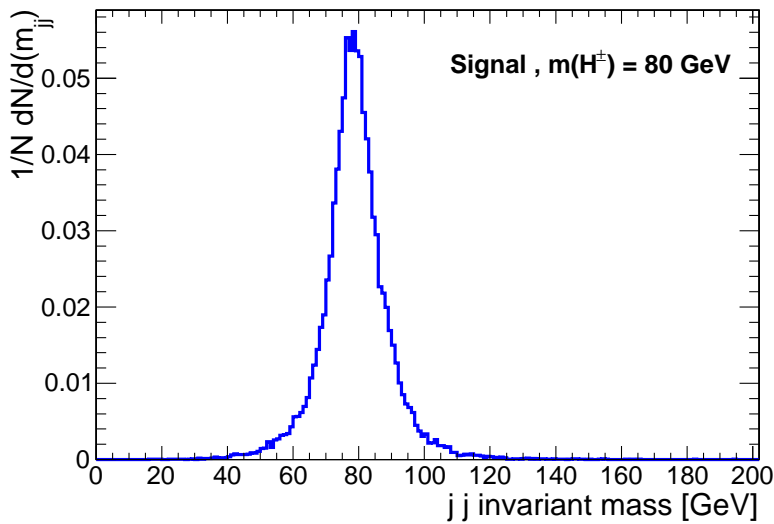}
\end{center}
\caption{The reconstructed W candidate invariant mass in signal sample with $m(H^{\pm})= 80 $ GeV.}
\label{wmass}
\end{figure}

\begin{figure}
\begin{center}
\includegraphics[width=0.6\textwidth]{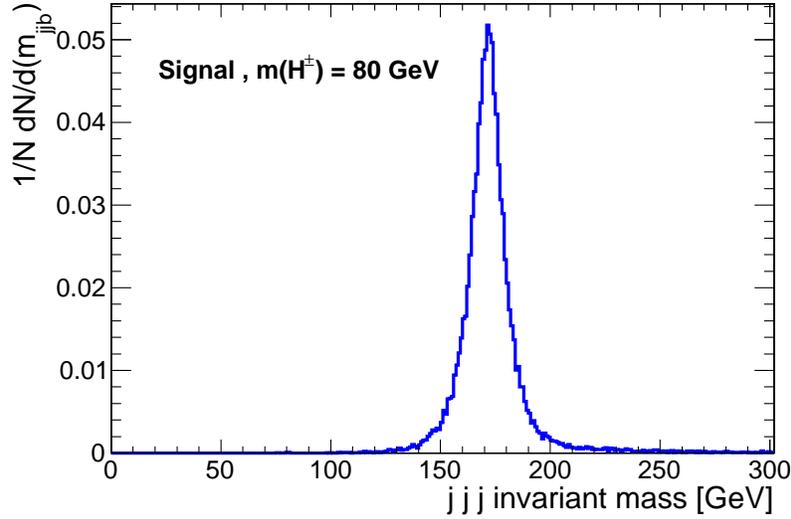}
\end{center}
\caption{The reconstructed top quark candidate invariant mass in signal sample with $m(H^{\pm})= 80 $ GeV.}
\label{topmass}
\end{figure}

\begin{figure}
\begin{center}
\includegraphics[width=0.6\textwidth]{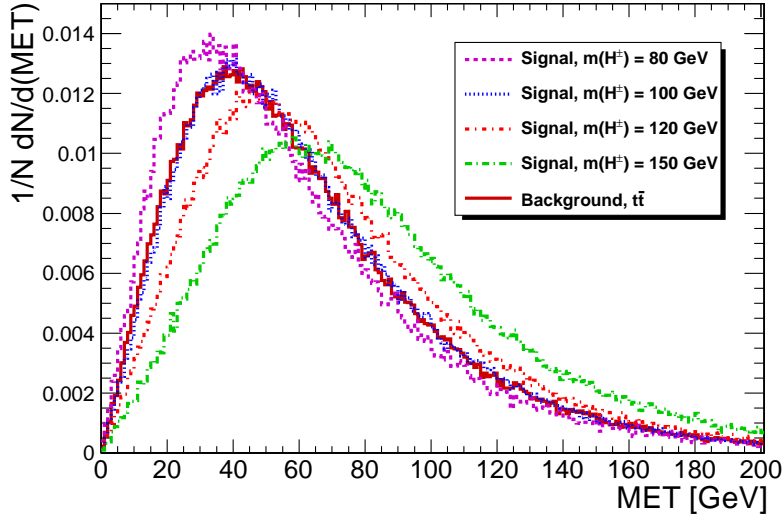}
\end{center}
\caption{Missing transverse energy distributions in signal events with different charged Higgs massed and the background sample.}
\label{met}
\end{figure}

\begin{figure}
\begin{center}
\includegraphics[width=0.6\textwidth]{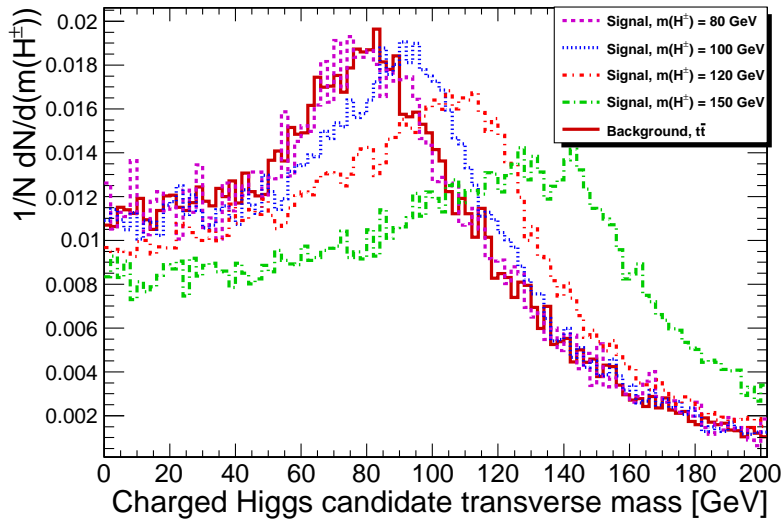}
\end{center}
\caption{The charged Higgs candidate transverse mass distribution in signal samples with different charged Higgs massed and the background sample.}
\label{chmass}
\end{figure}

In this section signal and background events are generated and their kinematic properties are compared and a set of selection cuts are introduced for the signal selection. Since the signal cross section is small, the analysis is designed for high luminosity regime of LHC. Therefore a total integrated luminosity of 300 $fb^{-1}$ is used to obtain the first estimate of the signal size.\\
Since signal events produce a muon, two light jets, two b-jets and some missing transverse energy through the following chain $t\bar{t} \ra H^{\pm}W^{\mp} b\bar{b} \ra \mu \nu j j b \bar{b}$, events can be triggered online by requiring a single muon trigger. Figure \ref{muonpt} shows the muon transverse momentum distributions in signal events with varying charged Higgs mass as well as the background distribution from the SM top pair production where the muon comes from the top decay to a W boson through $t\bar{t} \ra W^{\pm}W^{\mp} b\bar{b} \ra \mu \nu j j b \bar{b}$.\\
This is a mass independent analysis, i.e. the signal selection cuts do not depend on the charged Higgs mass. A more detailed study by variating selection cuts as a funcation of the charged Higgs mass may produce higher signal significances. However that requires a different study including cuts optimization which can be performed in a separated study. In the following a common set of selection cuts are applied on all charged Higgs mass cases. The cut on the muon $p_{T}$ and $|\eta|$ is applied as the following:\\
\be
\textnormal{Muon}~ p_{T} > 30~ \textnormal{GeV}, ~~~~ |\eta|<2.5 
\label{mupt}
\ee
The above requirement selects muons in the central detector region with the momentum threshold above 30 GeV. Harder cuts are avoided due to the small cross section of the signal, and the fact that they do not improve the overall significance of the signal. This was verified by evaluating $\epsilon^{i}_{S}/\sqrt{\epsilon^{i}_{B}}$ where $\epsilon^{i}_{S}$($\epsilon^{i}_{B}$) is the signal (background) relative efficiency when the i$th$ cut is applied. This ratio is a measure of goodness of a cut when the signal significance is going to be calculated as $S/\sqrt{B}$.
The next step starts with counting number of jets in the central region of $|\eta|<2.5$ and the jet transverse energy higher than 30 GeV. Softer jets (with lower $E_{T}$) are not taken into account to avoid reconstruction inefficiency and jet energy correction problems which happens for soft or forward-backward jets. In addition initial and final state radiation effect is reduced by this cut. All jets have to be separated enough from the muon in the event by requiring \\
\be
\Delta R_{(\textnormal{jet,muon})}>0.5
\label{drjm}
\ee
Here $\Delta R = \sqrt{(\Delta \eta)^2 + (\Delta \phi)^2}$ and $\phi$ is the azimuthal angle in the transverse plane.
Figure \ref{njet} shows the number of reconstructed jets which passed the $E_{T}$ and $|\eta|$ requirements mentioned above. Although this may not be a strongly distinguishing distribution, a cut on the number of jets is applied as the following:\\
\be
\textnormal{Number of jets} \geq 3, \textnormal{with} ~ E^{\textnormal{jet}}_{T} > 30~ \textnormal{GeV}, ~|\eta^{\textnormal{jet}}|<2.5 
\label{nj}
\ee
This cut ensures that events with low jet multiplicity (like single W events accompanied by 0, 1, 2 or more jets) 
do not pass the signal selection cuts. As is seen from Fig. \ref{njet}, the number of reconstructed jets has a peak at three which is the result of reconstruction inefficiency and the cuts applied
 on the jet transverse energy and pseudorapidity. Appearance of events with more than four jets in Fig. \ref{njet} is a consequence of initial and final state radiation (ISR and FSR). They contribute to only $\sim14\%$ of all events of the background and the signal with $m(H^{\pm}) = 80~$GeV. The estimation of the error on the number of such events requires a reasonable understanding of the ISR and FSR performance. The additional jets may also be produced by experimental effects such as electronic noise or detector material activation and a detailed study of all such effects needs a full simulation of events at the presence of detector effects which is beyond the scope of this analysis. However as a reference, we mention an analysis of $t\bar{t}$ events in the lepton+jet final state which is similar to the final state studied in this work. This analysis, reported in \cite{isr}, uses different PYTHIA samples generated by varying parameters responsible for the amount of ISR and FSR radiation based on the approach described in \cite{ATLASTDR}. The conclusion is summarized in small uncertainties of about +1.7$\%$ and -1.3$\%$ on the measured $t\bar{t}$ cross section depending on increasing or decreasing the ISR and FSR effects. This uncertainty has been obtained for $\sqrt{s}~=~7~$ TeV and should be evaluated for $\sqrt{s}~=~14~$ TeV, however, the overall conclusion is that we expect the uncertainty due to the extra jets associated with the $t\bar{t}$ pair to be at the level of few percent and it may not have a sizable effect on the signal significance. In the next step it is required that all jets should be separated by the $\Delta R$ requirement as the following:\\
\be
\Delta R_{(\textnormal{jet$_{i}$},\textnormal{jet$_{j}$})}>0.5
\label{drjj}
\ee
The indices $i$ and $j$ run over all selected jets in the event. Since at this point events contain three or more selected jets, this requirement makes sure that all pairs of those jets in the event are well separated and there is no overlap between the jet cones and they are well defined.\\
The number of reconstructed and selected jets are reduced with heavier charged Higgs in the event as the energy of the b-jet produced in the top quark decay to
 charged Higgs is suppressed when a heavy charged Higgs is produced. Therefore the b-jet passes
 the kinematic cuts with a lower efficiency resulting in a lower jet multiplicity. This feature 
can also be verified by counting number of b-tagged jets which are identified if the reconstructed jet 
lies within a cone of $\Delta R < 0.3$ around the parton level b-quark. An identification efficiency of 50$\%$ is also used when a jet matches a parton level b-quark. Figure \ref{nbjet} shows the number of b-tagged jets with the above algorithm.
The cut on the number of b-tagged jets defined as:\\
\be
\textnormal{Number of b-tagged jets} \geq 1, \textnormal{with} ~ E^{\textnormal{b-tagged jet}}_{T} > 30~ \textnormal{GeV}, ~|\eta^{\textnormal{b-tagged jet}}|<2.5 
\label{nbj}
\ee
not only selects $t\bar{t}$ events, but also rejects W+njets events which are produced with light jets, and also $W^{+}W^{-}$ and $ZZ$ events. The QCD multi-jet events are expected to be suppressed by this cut as well as the lepton requirement. \\
In the next step, as there is a W boson in the event which decays to two light jets, a $\chi^{2}$ minimization procedure is performed to reconstruct the W boson and the top quark masses.
The minimization is done by performing a loop over the jets in the event and finding the correct combination by minimizing the following $\chi^{2}$:\\
\be
\chi^{2}=(\frac{m_{j_l j_m}-m_{W}}{\sigma_{W}})^{2}+(\frac{m_{j_l j_m j_n}-m_{top}}{\sigma_{top}})^{2}
\label{chi}
\ee
where the loop is over the jet indices $l,~m$ and $n$; $m_{W}$ and $m_{top}$ are the nominal values of the W boson and the top quark masses, i.e. 80 and 175 GeV respectively and the $\sigma_{W}$ and $\sigma_{top}$ values are taken to be 10 and 15 GeV respectively. Thanks to the kinematic cuts on jets this process is easily performed as every event has few hard jets to combine. Figures \ref{wmass} and \ref{topmass} show the reconstructed masses of W boson and top quark in signal events with $m(H^{\pm})= 80 $ GeV. All other signal samples and the background sample show very similar distributions as the hadronic sector of these events, i.e. $t \ra W^{+}b \ra jjb$ is identical.\\
To ensure that the selected event has a top quark and W boson candidate in the hadronic sector, a mass window cut is applied as the following:\\
\be
|m_{jj}-m_{W}|<10~ \textnormal{GeV},~~|m_{jjj}-m_{top}|<20~ \textnormal{GeV}
\label{mjj}
\ee
It should be noted that here no b-tagging has been used for the top reconstruction although in the previous step the b-jet requirement has already been applied. \\
Finally the missing transverse energy is calculated as the negative sum over momentum components of all stable particles, which lie in the region $|\eta|<3.5$, exclusing neutrinos in the event. A more general calculation should exclude sneutrinos, neutralinos, gravitons, gravitinos, ..., however in this analysis the relevant sources of missing $E_{T}$ are neutrinos. Of course in the real situation, this quantity is calculated from the imbalance of energy in the transverse plane of the detector. Figure \ref{met} shows distributions of missing $E_{T}$ in signal and background. 
A cut on the missing $E_{T}$ in the event is applied as the following :\\
\be
\textnormal{Missing}~ E_{T} > 30~ \textnormal{GeV}
\label{meteq}
\ee
As is seen from Fig. \ref{met}, distributions are similar producing no sizable increase in the signal significance when harder cuts are applied.\\
Another quantity which can be made is the charged Higgs candidate transverse mass. This is calculated as the following:\\
\be
m_{T}=\sqrt{2.p^{\textnormal{muon}}_{T}.E^{\textnormal{miss}}_{T}.(1-cos\Delta\phi_{(\textnormal{muon},E^{\textnormal{miss}}_{T})})}
\ee
where $\Delta\phi_{(\textnormal{muon},E^{\textnormal{miss}}_{T})}$ is the angle between the muon flight and the missing $E_{T}$ in the transverse plane. Figure \ref{chmass} shows the distribution of this quantity in signal samples and the background. 
The lower cut applied on the charged Higgs transverse mass is set as the following:\\
\be
\textnormal{Charged Higgs candidate}~ m_{T}>50~ GeV
\label{cheq}
\ee
As the signal to background ratio is small, harder cuts end up with smaller statistics while not increasing the signal significance sizably. Now the signal and background relative and total efficiencies are calculated to evaluate the signal significance when all cuts are applied. Table \ref{sigsel} shows the signal selection efficiencies for different charged Higgs masses. Table \ref{bsel} shows the background efficiencies and the final event yield.
\begin{table}
\begin{center}
\begin{tabular}{|c|c|c|c|c|}
\hline
$m_{(H^{\pm})}$ & 80 GeV  & 100 GeV & 120 GeV & 150 GeV \\
\hline
$\sigma \times$ BR & 302 fb & 228 fb & 149 fb & 38 fb \\
\hline
Muon selection (Eq. \ref{mupt}) & 72.1 $\%$  & 79.6 $\%$& 85.5 $\%$& 91$\%$ \\
\hline
Number of jets $\geq$ 3 (Eq. \ref{nj})& 72.4 $\%$ & 70.2 $\%$& 65.9 $\%$& 52.5 $\%$\\
\hline
$\Delta R(j,j)>$ 0.5 (Eq. \ref{drjj})& 87.6$\%$ & 87.2$\%$ & 87$\%$ & 88$\%$ \\
\hline 
Number of b-jets $\geq$ 1 (Eq. \ref{nbj})& 62.7 $\%$ & 61.9 $\%$& 59.7 $\%$& 49.7$\%$ \\
\hline
W and top mass window (Eq. \ref{mjj})& 79.1 $\%$& 79.2 $\%$& 79.9$\%$ & 80.6$\%$ \\
\hline
MET $>$ 30 GeV (Eq. \ref{meteq})& 69.1 $\%$& 78.4 $\%$& 85$\%$ & 91.2$\%$ \\
\hline
Charged Higgs mass window (Eq. \ref{cheq})& 70.7 $\%$& 71.8$\%$ & 74.4$\%$ & 78.9 $\%$\\
\hline
Total efficiency & 11.1 $\%$& 13.47$\%$ & 14.8 $\%$& 12.1$\%$ \\
\hline
Events @ 300 $fb^{-1}$ & 10057 & 9213 & 6616 & 1379\\
\hline
\end{tabular}
\end{center}
\caption{Signal selection efficiencies.\label{sigsel}}
\end{table}

\begin{table}
\begin{center}
\begin{tabular}{|c|c|}
\hline
$\sigma \times$ BR & 124000 fb \\
\hline
Muon selection (Eq. \ref{mupt}) & 69.5$\%$ \\
\hline
Number of jets $\geq$ 3 (Eq. \ref{nj})& 72.5 $\%$\\
\hline
$\Delta R(j,j)>$ 0.5 (Eq. \ref{drjj})& 87.6 $\%$\\
\hline 
Number of b-jets $\geq$ 1 (Eq. \ref{nbj})& 62.4 $\%$\\
\hline
W and top mass window (Eq. \ref{mjj})& 79.7$\%$ \\
\hline
MET $>$ 30 GeV (Eq. \ref{meteq})& 74.5 $\%$\\
\hline
Charged Higgs mass window (Eq. \ref{cheq})& 70.7 $\%$\\
\hline
Total efficiency & 11.6 $\%$\\
\hline
Events @ 300 $fb^{-1}$ & 4315000 \\
\hline
\end{tabular}
\end{center}
\caption{$t\bar{t}$ background selection efficiencies.\label{bsel}}
\end{table}
The numbers quoted in Tabs. \ref{sigsel} and \ref{bsel} lead to the following results listed in Tab. \ref{significance}. As is seen the signal significance is close to 5$\sigma$ near the direct search lower limit. Although the signal efficiency increase with higher charged Higgs masses which is due to the harder kinematics of events, the signal significance decreases due to the reduction in cross section. A problem to note is that the signal to background ratio is very small making the statistical conclusions about the existence of the signal, a challenging task.
\begin{table}
\begin{center}
\begin{tabular}{|c|c|c|c|c|}
\hline
$m(H^{\pm})$ &  80 GeV & 100 GeV & 120 GeV & 150 GeV \\
\hline
$S/B$ & 0.002 & 0.002 & 0.001 & 0.0003 \\
\hline
$S/\sqrt{B}$ & 4.8 & 4.4 & 3.2 & 0.66 \\
\hline
\end{tabular}
\end{center}
\caption{Signal to Background ratio and the signal significance.\label{significance}}
\end{table}

\section{Discussion on the Accuracy and Consistency of Results}
The analysis presented here does not reflect completely the LHC experimental environment. A detailed study of this channel requires a full simulation of an LHC detector including all experimental environment in detail followed by a full analysis of experimental and theoretical uncertainties involved in the analysis which is beyond the scope of this paper. This can be done by one of LHC experiments in future with access to the full set of a real analysis software. In this work, however, a result consistency check is performed between different software versions. \\
The PYTHIA 6.4.21 used for the analysis is one of the most recent versions of the $6^{th}$ series of this package. Although PYTHIA 8 series are taking over, PYTHIA 6 is still being used by LHC experiments in their physics simulations. Therefore we rely on PYTHIA 6 results as the basis. \\
Concerning HDECAY, the up-to-date version 4.1 was compared with the version used in the analysis, i.e., version 3.5. No sizable difference between the branching ratios given by the two above versions was observed.\\
The MCFM version 5.7 which was used in the analysis was also compared with the up-to-date version 6.1. In this case also, no change in the cross sections was observed up to the femtobarn level.\\
The choice of the PDF set can potentially have some impact on the event cross section as well as selection efficiencies. This was verified by using LHAPDF 5.8.6 (the up-to-date version) and using three sets of PDFs, i.e., MSTW (2008), CTEQ6mE (2002) and NNPDF 2.1 with 100 fit parameters (2011). The MSTW and MRST package used in the analysis calculate $\alpha_{s}$ evolution up to NNLO in its perturbation series, while CTEQ6mE and NNPDF 2.1 evolve $\alpha_{s}$ up to NLO. Results for signal and backgrounds are listed in Tabs. \ref{pdf1} and \ref{pdf2}. As seen from these tables, the overall change in the final results in terms of signal significance is expected to be at the level of few percent. There can be more effects due to $\alpha_{s}$ variations in PDFs which was not studied in this analysis, however, the overall effect of PDF+$\alpha_{s}$ uncertainty in the results is expected to remain at the level of few percent. An example of such analyses, although done for a different process, can be found in \cite{pdf}. \\
Finally, consistency of the results can be verified through Figs. \ref{muonpt},\ref{njet},\ref{nbjet},\ref{met} and \ref{chmass} which reveal a general effect of increasing the charged Higgs mass which produces harder events with higher muon $p_{T}$, missing transverse energy, etc. On the contrary the heavier the charged Higgs, the less number of jets identified. This is due to the fact that the $b$-jet produced in the top quark decay is suppressed when increasing the charged Higgs mass resulting in a softer jet which is identified with a smaller probability as it may not pass the transverse energy threshold applied in the analysis. 
\begin{table}
\begin{center}
\begin{tabular}{|c|c|c|}
\hline
PDF set & efficiency & Cross section \\
\hline
MSTW 2008 & -2$\%$ & 0.2$\%$ \\
\hline
CTEQ6mE & -0.3$\%$ & -2$\%$ \\
\hline
NNPDF2.1 & 4$\%$ & 2$\%$ \\
\hline
\end{tabular}
\end{center}
\caption{Relative change in the total efficiency and cross section of the signal compared to the PDF set used in the analysis (MRST 2006) \label{pdf1}}
\end{table}
\begin{table}
\begin{center}
\begin{tabular}{|c|c|c|}
\hline
PDF set & efficiency & Cross section \\
\hline
MSTW 2008 & 4$\%$ & -0.1$\%$ \\
\hline
CTEQ6mE & -1$\%$ & -1$\%$ \\
\hline
NNPDF2.1 & 1$\%$ & 4$\%$ \\
\hline
\end{tabular}
\end{center}
\caption{Relative change in the total efficiency and cross section of the background compared to the PDF set used in the analysis (MRST 2006) \label{pdf2}}
\end{table}
   
\section{Conclusions}
The charged Higgs decay to muon and muon neutrino was studied in this work. Results show that the light charged Higgs may in fact be observable through this decay channel. The signal significance ranges between 4.8$\sigma$ to 0.66$\sigma$ for 80 GeV $< m(H^{\pm}) <$ 150 GeV with \tanb = 20. The signal yield is few thousand events, however, a challenging problem would be the very Small S/B ratio which is of the order of $10^{-3}$ or smaller. The singal significance decreases rapidly with increasing the charged Higgs mass, however, the higher charged Higgs mass, the smaller the excluded area in ($m(H^{\pm}),$ \tanb) plane. As an example, for $m(H^{\pm}) \geq$ 150 GeV, \tanb values are not excluded up to \tanb $\simeq$ 50. This point in the parameter space leads to a signal significance of 3.7$\sigma$. Although other charged Higgs decay channels produce higher signal rates, their 5$\sigma$ contours do not cover the parameter space completely. Therefore adding this channel with a proper statistical method may increase the overall sensitivity of LHC detectors to the charged Higgs signal. 
\pagebreak

\end{document}